\documentclass[11pt,a4paper]{article}

\pdfoutput=1

\usepackage{jheppub}
\usepackage{amsmath,amsfonts,mathrsfs,graphicx,bbm}
\usepackage{fixltx2e}
\usepackage{calligra}

\usepackage[font=small]{caption}

\usepackage{lmodern}

\usepackage{graphicx}
\usepackage{booktabs}
\usepackage{dsfont}

\usepackage{mathrsfs}
\usepackage{amsmath,amssymb}
\usepackage{mathtools}
\usepackage{bbm}
\usepackage{slashed}
\usepackage{braket}

\usepackage{hyperref}

\usepackage{xspace}

\usepackage{color}

\definecolor{grey}{rgb}{0.4,0.4,0.5}
\definecolor{darkgreen}{rgb}{0,0.5,0}
\definecolor{darkred}{rgb}{0.6,0.0,0}
\definecolor{lightbrown}{rgb}{1,0.9,0.8}
\definecolor{brown}{rgb}{0.6,0.3,0.3}
\definecolor{darkblue}{rgb}{0,0,0.8}
\definecolor{darkmagenta}{rgb}{0.5,0,0.5}

\newcommand{\mO}{\mathcal{O}}

\newcommand{\mN}{\mathcal{N}}

\numberwithin{equation}{section}

\makeatletter
 \let\old@startsection=\@startsection
 \let\oldl@section=\l@section
 \renewcommand{\@startsection}[6]{\old@startsection{#1}{#2}{#3}{#4}{#5}{#6\mathversion{bold}}}
 \renewcommand{\l@section}[2]{\oldl@section{\mathversion{bold}#1}{#2}}
\makeatother

\renewcommand{\leq}{\leqslant}
\renewcommand{\geq}{\geqslant}

\DeclareMathAlphabet{\mathcalligra}{T1}{calligra}{m}{n}

\def\XXint#1#2#3{{\setbox0=\hbox{$#1{#2#3}{\int}$}
    \vcenter{\hbox{$#2#3$}}\kern-.5\wd0}}

\newcommand{\alg}[1]{\mathfrak{#1}}

\newcommand{\be}{\begin{equation}}
\newcommand{\ee}{\end{equation}}
\newcommand{\bea}{\begin{eqnarray}}
\newcommand{\eea}{\end{eqnarray}}
\newcommand{\bal}{\begin{equation}\begin{aligned}}
\newcommand{\eal}{\end{aligned}\end{equation}}

\newcommand{\bee}{\begin{enumerate}}
\newcommand{\eee}{\end{enumerate}}
\newcommand{\bei}{\begin{itemize}}
\newcommand{\eei}{\end{itemize}}

\newcommand{\ads}{${\rm  AdS}_5\times {\rm S}^5\ $}

\def\ov{\over}

\def\s {\sigma}

\def\p{\phi}

\def\m {\mu}

\def\x'{\mathaccent 19 x}
\def\y'{\mathaccent 19 y}
\def\n'{\mathaccent 19 n}
\def\u'{\mathaccent 19 u}

\def\et'{\mathaccent 19 \eta}
\def\th'{\mathaccent 19 \theta}
\def\lam'{\mathaccent 19 \lambda}
\def\varet'{\mathaccent 19 \vartheta}
\def\rh'{\mathaccent 19 \rho}
\def\ph'{\mathaccent 19 \phi}
\def\xb'{\mathaccent 19 {\bar{x}}}

\def\sl(2){\alg{sl}(2)}

\def\be{\begin{equation}}
\def\ee{\end{equation}}

\def\s {\sigma}

\def\p{\phi}

\def\ov{\over}

\def\d1{{\dot{1}}}

\newcommand{\bem}{\left (\begin{matrix}}
\newcommand{\eem}{\end{matrix} \right )}

\usepackage{mciteplus}
\title{Four-point functions of 1/2-BPS operators of any weights in the supergravity approximation}

\author[a,1]{Gleb Arutyunov}
\note{Correspondent fellow at Steklov
Mathematical Institute, Moscow.}
\author[a]{Rob Klabbers}
\author[a]{and Sergei Savin}
\affiliation[a]{II. Institut f\"ur Theoretische Physik, Universit\"at Hamburg, Luruper Chaussee 149, 22761 Hamburg, Germany\\
Zentrum f\"ur Mathematische Physik, Universit\"at Hamburg, Bundesstrasse 55, 20146 Hamburg, Germany
}
\emailAdd{gleb.arutyunov@desy.de}  
\emailAdd {rob.klabbers@desy.de}
\emailAdd{sergei.savin@desy.de}

\abstract{We present the computation of all the correlators of 1/2-BPS operators in $\mN=4$ SYM with weights up to $8$ as well as some very high-weight correlation functions from the effective supergravity action. The computation is done by implementing the recently developed simplified algorithm in combination with the harmonic polynomial formalism. We provide a database of
these results attached to this publication and additionally check for
almost all of the functions in this database that they agree with the
conjecture on their Mellin-space form.
}

\begin{document}
\begin{flushright}
\scriptsize{[ZMP-HH/18-15]}
\end{flushright}

\maketitle
\flushbottom

\renewcommand{\thefootnote}{\arabic{footnote}}
\setcounter{footnote}{0}

\section{Introduction}
Recently the problem of computing the four-point functions of 1/2-BPS operators in planar $\mN=4$ SYM at strong coupling has received new attention, due to the appearance of a surprisingly simple formula for the correlators of arbitrary weights in the Mellin space, conjectured by the authors of \cite{Rastelli:2016nze, Rastelli:2017udc}. This conjecture, based on symmetry and physical assumptions, matched all the examples known then in the literature \cite{DHoker:1999kzh, Arutyunov:2000py, Arutyunov:2002fh, Arutyunov:2003ae, Berdichevsky:2007xd, Uruchurtu:2008kp, Uruchurtu:2011wh}, including the infinite family studied in \cite{Uruchurtu:2011wh}.  However, the fact that all these examples are in one way or another degenerate has required more tests. The first step in this direction was to prove that one of the assumptions on which the work \cite{Rastelli:2016nze,Rastelli:2017udc} was based, namely that the four-derivative Lagrangian in the effective supergravity action vanishes in general, holds. This statement, which was already observed in all the concrete examples \cite{DHoker:1999kzh, Arutyunov:2000py, Arutyunov:2002fh, Arutyunov:2003ae, Berdichevsky:2007xd, Uruchurtu:2008kp, Uruchurtu:2011wh}, was recently proved in \cite{Arutyunov:2017dti}. The next test was to check the conjecture on less degenerate cases, which was recently initiated in \citep{Arutyunov:2018neq} by computing the $\langle 2345 \rangle$ and $\langle 3456 \rangle$ correlators. These functions consist of all-different weight operators and are far from the extremality condition. It was shown that these correlators perfectly match with the Mellin formula. Apart from that, the algorithm of computing supergravity correlators from the compactified string action was significantly simplified, opening the possibility for calculating more complicated cases. 

The knowledge of more complicated four-point functions could be used to further test the Mellin conjecture \cite{Rastelli:2016nze, Rastelli:2017udc}, but also to probe the non-planar spectrum of $\mN=4$ SYM by computing $1/N$ corrections around the supergravity correlator \cite{Alday:2017xua,Aprile:2017xsp,Aprile:2017qoy,Aprile:2018efk}. 

In this work we combine the simplifications, obtained in \citep{Arutyunov:2018neq}, with the harmonic polynomial formalism \cite{Dolan:2003hv,Nirschl:2004pa}. The latter formalism provides a further significant simplification of the computational algorithm, allowing us to find the four-point functions of CPOs for practically any given weights. We discuss the explicit computation of all the correlation functions with weights up to $8$ as well as $\langle 7\,10\,12\,17 \rangle$ and $\langle 17 \, 21 \, 23 \, 25 \rangle$. This is a major improvement over the previously available set of correlators. We furthermore check whether these correlators match the Mellin conjecture. We make all the results available in a database attached to this publication.

This paper, which follows the notation from \cite{Arutyunov:2018neq}, is organized as follows: in section \ref{sec:typeIIB} we recall how to compute four-point correlation functions from the type-IIB supergravity effective action. In section \ref{sec:harmonicpolynomial} we discuss how one can use the harmonic polynomial formalism to simplify the computation of the $a$, $p$ and $t$ tensors. In section \ref{sec:results} we discuss how we use this to compute new correlation functions. In particular, we discuss that all these new examples match the Mellin conjecture from \cite{Rastelli:2016nze,Rastelli:2017udc}. In section \ref{sec:conclusions} we conclude. In the appendix we discuss a way to obtain coordinate-space four-point functions from their Mellin-space form. 
\section{Four-point functions from type IIB supergravity}
\label{sec:typeIIB}
The central objects of our study are the four-point correlation functions in planar $\mN=4$ SYM theory in the strong coupling limit\footnote{We follow the notation from \cite{Arutyunov:2018neq}.}
\be
\label{eq:correlator}
\langle k_1 k_2 k_3 k_4 \rangle \coloneqq \langle \tilde{\mathcal{O}}_{k_1}\left(x_1,t_1 \right)\cdots\tilde{\mathcal{O}}_{k_4}\left(x_4,t_4 \right) \rangle,
\ee
where the $x_i$ are the space-time coordinates and the $t$'s are six-dimensional null vectors introduced to keep track of the R-symmetry. Here, the operators
\be
\label{eq:BPSdecomp}
\tilde{\mO}_k = \tilde{\mO}^{i_1\ldots i_k} t_{i_1} \cdots t_{i_k}
\ee
are $1/2$-BPS operators of conformal weight $k$, transforming in the representation of the SO$(6)$ group specified by the Dynkin labels $[0,k,0]$, and are in general mixtures of single- and double-trace operators \cite{Arutyunov:2000ima}\footnote{See section $7$ of \cite{Arutyunov:2018neq} for a more elaborate discussion using the notation used in this paper.}:
\be
\tilde{\mO}^{i_1,\ldots , i_k} = \kappa_k \text{Tr}\left( \phi^{i_1}\ldots \phi^{i_k}\right) -
\sum_{\substack{k_1,k_2\geq 2 \\ k_1 + k_2 =k}} \frac{\lambda_{k,k_1,k_2}}{2N}\kappa_{k_1}\kappa_{k_2} \text{Tr}\left( \phi^{i_1}\ldots \phi^{i_{k_1}}\right) \text{Tr}\left( \phi^{i_{k_1+1}}\ldots \phi^{i_k}\right),
\ee 
where the $\phi$ are the scalar operators of $\mN=4$ SYM, the $\kappa_k$ are normalization constants, the $\lambda_{k,k_1,k_2}$ are fixed real numbers and we omit the total symmetrization over indices as this gets taken care of in the contraction with $t$ vectors in \eqref{eq:BPSdecomp}.

According to the AdS/CFT correspondence \cite{Maldacena:1997re, Witten:1998qj, Gubser:1998bc} these operators are dual to the Kaluza-Klein modes of compactified type-IIB supergravity on \ads. When computing the four-point function of such operators, the boundary values of the Kaluza-Klein modes act as sources with the generating functional being given by the on-shell value of the string partition function $\exp(-S_{IIB})$. This approach warrants knowledge of the supergravity action up to fourth order in the fields, which was obtained in \cite{Arutyunov:1998hf, Arutyunov:1999en, Arutyunov:1999fb}. In these works explicit expressions were obtained for the coupling constants in the effective action in terms of so-called $a$, $p$ and $t$ tensors which are effectively Clebsch-Gordan coefficients for the representations of SO$(6)$. Nevertheless, since the action, which splits into a \emph{contact} and an \emph{exchange} part, is still very complicated and finding the $a$, $p$ and $t$ tensors requires additional work executing the algorithm to compute four-point functions from this action is far from trivial. 

Recently in \cite{Arutyunov:2018neq} significant simplifications of this algorithm were obtained: the first simplification makes it possible to bypass having to write the full Lagrangian down, which becomes unpleasant very fast because of the growing number of descendants coupled to scalars in the cubic terms. Instead, the streamlined procedure allows one to obtain the four-point function for a given set of weights straight away. For example, the contribution to the \textit{non-normalized} correlator, coming from the exchange terms, can be written as the following sum of the $s$, $t$ and $u$ channels:
\small{
\bea
\label{eq:LExchange}
	\langle k_1 k_2 k_3 k_4 \rangle_{\text{Exchange}}&=&
	\left\{\sum {36\ov \zeta(s_k)}S^{I_1I_2I_k}S^{I_3I_4I_k}\mathbf{S}^k_{k_1k_2k_3k_4} + \sum {4\ov \zeta(t_k)}T^{I_1I_2I_k}T^{I_3I_4I_k}\mathbf{S}^{k+4}_{k_1k_2k_3k_4}\right.\nonumber\\
	&+& \sum {4\ov \zeta(\p_k)}\Phi^{I_1I_2I_k}\Phi^{I_3I_4I_k}\mathbf{S}^{k+2}_{k_1k_2k_3k_4}
	+ \sum{1\ov \zeta(A_{\m,k})}A^{I_1I_2I_k}A^{I_3I_4I_k}\mathbf{V}^k_{k_1k_2k_3k_4}\nonumber\\
	&+& \left. \sum{1\ov \zeta(C_{\m,k})}C^{I_1I_2I_k}C^{I_3I_4I_k}\mathbf{V}^{k+2}_{k_1k_2k_3k_4} + \sum{4\ov\zeta(\varphi_{\m\nu,k})}G^{I_1I_2I_k}G^{I_3I_4I_k}\mathbf{T}^k_{k_1k_2k_3k_4} \right\}_{\text{s}} \nonumber\\
	&&~ \nonumber	\\
	&+&\left\{\sum {36\ov \zeta(s_k)}S^{I_1I_3I_k}S^{I_2I_4I_k}\mathbf{S}^k_{k_1k_3k_2k_4} + \sum {4\ov \zeta(t_k)}T^{I_1I_3I_k}T^{I_2I_4I_k}\mathbf{S}^{k+4}_{k_1k_3k_2k_4}\right.\nonumber\\
	&+& \sum {4\ov \zeta(\p_k)}\Phi^{I_1I_3I_k}\Phi^{I_2I_4I_k}\mathbf{S}^{k+2}_{k_1k_3k_2k_4}
	+ \sum{1\ov \zeta(A_{\m,k})}A^{I_1I_3I_k}A^{I_2I_4I_k}\mathbf{V}^k_{k_1k_3k_2k_4}\nonumber\\
	&+& \left. \sum{1\ov \zeta(C_{\m,k})}C^{I_1I_3I_k}C^{I_2I_4I_k}\mathbf{V}^{k+2}_{k_1k_3k_2k_4} + \sum{4\ov\zeta(\varphi_{\m\nu,k})}G^{I_1I_3I_k}G^{I_2I_4I_k}\mathbf{T}^k_{k_1k_3k_2k_4} \right\}_{\text{t}} \nonumber\\
	&&~ \nonumber	\\
	&+&\left\{\sum {36\ov \zeta(s_k)}S^{I_1I_4I_k}S^{I_2I_3I_k}\mathbf{S}^k_{k_1k_4k_2k_3} + \sum {4\ov \zeta(t_k)}T^{I_1I_4I_k}T^{I_2I_3I_k}\mathbf{S}^{k+4}_{k_1k_4k_2k_3}\right.\nonumber\\
	&+& \sum {4\ov \zeta(\p_k)}\Phi^{I_1I_4I_k}\Phi^{I_2I_3I_k}\mathbf{S}^{k+2}_{k_1k_4k_2k_3}
	+ \sum{1\ov \zeta(A_{\m,k})}A^{I_1I_4I_k}A^{I_2I_3I_k}\mathbf{V}^k_{k_1k_4k_2k_3}\nonumber\\
	&+& \left. \sum{1\ov \zeta(C_{\m,k})}C^{I_1I_4I_k}C^{I_2I_3I_k}\mathbf{V}^{k+2}_{k_1k_4k_2k_3} + \sum{4\ov\zeta(\varphi_{\m\nu,k})}G^{I_1I_4I_k}G^{I_2I_3I_k}\mathbf{T}^k_{k_1k_4k_2k_3} \right\}_{\text{u}} \nonumber\\
\eea
}
Here the exchange Witten diagrams, $\mathbf{S}$, $\mathbf{V}$ and $\mathbf{T}$ are multiplied by the corresponding combination of quadratic couplings $\zeta$ and cubic couplings $S,T,\Phi,A,C$ and $G$ that were derived in \cite{Arutyunov:1998hf, Arutyunov:1999en, Arutyunov:1999fb}. The exchange Witten diagrams can be expressed in terms of exchange integrals. A simple method to compute them was developed in \citep{DHoker:1999mqo} and further generalizations appeared in \citep{Arutyunov:2002fh} and \cite{Berdichevsky:2007xd}. The appearing sums run over the possible set of exchange fields in the relevant channel restricted by the SU$(4)$ selection rule. 

Finding the contact part also forms no difficulty once all the quartic couplings are computed, as was discussed in \cite{Arutyunov:2018neq}. Therefore the main remaining difficulty in computing the correlator sits in finding the $a$, $p$ and $t$ tensors that form the building blocks of the couplings. Their explicit expressions depend on the chosen weights and their computation becomes complicated quite quickly. In \cite{Arutyunov:2018neq} we discussed how one can simplify the procedure outlined in \cite{Arutyunov:2002fh} somewhat by carefully analyzing the sums over the symmetric group necessary in doing the tensor contractions. 

\section{Harmonic polynomial formalism}
\label{sec:harmonicpolynomial}
However, the computation of the $a$, $p$ and $t$ tensors simplifies incredibly in the \textit{harmonic polynomial} formalism, developed in \cite{Dolan:2003hv,Nirschl:2004pa} and applied to compute some supergravity correlators in \cite{Uruchurtu:2008kp, Uruchurtu:2011wh}. It turns out that, after an appropriate normalization, the $a$, $p$ and $t$ tensors can be expressed as harmonic polynomials $Y_{nm}^{(a,b)}(\s,\tau)$ in the new variables $\sigma$ and $\tau$ 
\bea
	\s\equiv\frac{t_{13}t_{24}}{t_{12}t_{34}} \qquad \text{and} \qquad \tau\equiv\frac{t_{14}t_{23}}{t_{12}t_{34}},\qquad \text{where} \qquad t_{ij}\equiv t_i\cdot t_j,
\eea
which carry the non-trivial dependence on the null vectors $t_i$. These functions are generalized eigenfunctions of the SO$(6)$ Casimir operator $L^2$, satisfying
\begin{equation}
L^2\left(t_{14}^a t_{24}^b Y_{nm}^{(a,b)}(\s,\tau) \right) = -2 C_{nm} t_{14}^a t_{24}^b Y_{nm}^{(a,b)}(\s,\tau),
\end{equation}
with $C_{nm}$ being the corresponding eigenvalue. Moreover, one can solve this equation \cite{Nirschl:2004pa} and find that the $Y_{nm}^{(a,b)}$ can be expressed explicitly in terms of Jacobi polynomials $P_{n}^{(a,b)}$:
\bea
	Y_{nm}^{(a,b)}(\s,\tau)=\frac{2 (n+1)! (a+b+n+1)!}{(a+1)_m (b+1)_m (a+b+2 n+2)!}P_{nm}^{(a,b)}(\s,\tau)\, ,
\eea
where $(\ldots)_m$ is the usual Pochhammer symbol and
\bea
	P_{nm}^{(a,b)}(y,\bar y)=\frac{P_{n+1}^{(a,b)}(y)P_{m}^{(a,b)}(\bar y)-P_{m}^{(a,b)}(y)P_{n+1}^{(a,b)}(\bar y)}{y-\bar y}\,,
\eea
which can be related to the original $\sigma$ and $\tau$ variables via 
\bea
	\sigma=\frac{1}{4}(y+1)(\bar{y}+1)\qquad \text{and} \qquad \tau=\frac{1}{4}(1-y)(1-\bar{y}).
\eea
It was discussed in \cite{Uruchurtu:2011wh}, that the product of $C$ tensors appearing in the product of scalar $a_{125}a_{345}$, vector $t_{125}t_{345}$ and tensor $p_{125}p_{345}$ harmonics for arbitrary weights with fixed exchange leg $k_5$ are proportional to these $Y_{nm}^{(a,b)}$:
\begin{equation}
\begin{aligned}
a_{125}a_{345} &\sim & \langle C^{I_1}_{k_1}C^{I_2}_{k_2}C^I_{[0,a+b+2m,0]}\rangle \langle C^{I_3}_{k_3}C^{I_4}_{k_4}C^I_{[0,a+b+2m,0]}\rangle & = & \mathcal{T}\mathcal{B}_a Y_{mm}^{(a,b)},\,\,\, \\
t_{125}t_{345} &\sim & \langle C^{I_1}_{k_1}C^{I_2}_{k_2}C^I_{[1,a+b+2m,1]}\rangle \langle C^{I_3}_{k_3}C^{I_4}_{k_4}C^I_{[1,a+b+2m,1]}\rangle & = & \mathcal{T}\mathcal{B}_t Y_{m+1,m}^{(a,b)} ,\\
p_{125}p_{345} &\sim &\langle C^{I_1}_{k_1}C^{I_2}_{k_2}C^I_{[2,a+b+2m,2]}\rangle \langle C^{I_3}_{k_3}C^{I_4}_{k_4}C^I_{[2,a+b+2m,2]}\rangle & = & \mathcal{T}\mathcal{B}_p Y_{m+2,m}^{(a,b)} ,
\end{aligned}
\end{equation}
where the $t$-dependent prefactor $\mathcal{T}$ is given by
\begin{equation}
\mathcal{T} = t_{12}^{k_3} t_{13}^{b} t_{14}^{a} t_{34}^{\frac{k_1 +k_2+k_3 -k_4}{2}}
\end{equation}
and $k_5$ satisfies
\begin{equation}
\label{eq:k5restriction}
k_5 = a+b+2m, \quad k_5 = a+b+2m+1 \quad\text{ or } \quad k_5 = a+b+2m+2
\end{equation}
for some nonnegative integer $m$ respectively. The proportionality coefficients $\mathcal{B}$ were worked out in \cite{Uruchurtu:2011wh}: given a set of weights $k_1,k_2,k_3,k_4$ ordered such that 
\begin{equation}
a = \frac{k_1 +k_4 - k_2 -k_3}{2}, \quad b = \frac{k_2 +k_4 - k_1 -k_3}{2}
\end{equation}
are nonnegative\footnote{This might require a shuffle $\{k_1,k_2,k_3,k_4 \}\rightarrow\{k_3,k_4,k_1,k_2 \}$, which is certainly possible in all cases.} and an intermediate weight $k_5$ satisfying \eqref{eq:k5restriction} they take the following form:
\begin{equation}
\begin{aligned}
\mathcal{B}_a =& \frac{\alpha_{251}!\alpha_{512}!\alpha_{453}!\alpha_{534}!}{a! b! k_5!}, \\
\mathcal{B}_t =& \frac{(k_5+1)(\alpha_{251}-\tfrac{1}{2})!(\alpha_{512}-\tfrac{1}{2})!(\alpha_{453}-\tfrac{1}{2})!(\alpha_{534}-\tfrac{1}{2})!}{a! b!k_5!}, \\
\mathcal{B}_p =& 2^4\cdot\frac{\alpha_{251}!\alpha_{512}!\alpha_{453}!\alpha_{534}!}{a! b! k_5!(k_5+1)!},
\end{aligned}
\end{equation}
where $\alpha_{123} = \tfrac{k_1+k_2-k_3}{2}$. This now allows for a straightforward evaluation of $a_{125}a_{345}$, $t_{125}t_{345}$ and $p_{125}p_{345}$ for any weights as all the complicated tensor structure is captured by Jacobi polynomials. To obtain the corresponding tensors in the $t$ and $u$ channel, e.g. $a_{135}a_{245}$ and $a_{145}a_{235}$, one simply reshuffles the $t_i$. 
\section{Results}
\label{sec:results}
\paragraph{Computation.} The aforementioned simplifications, obtained in \cite{Arutyunov:2018neq}, in combination with the harmonic polynomial formalism reviewed in the previous section allow one to compute any supergravity four-point function of the CPOs in \eqref{eq:correlator} of given weights in very little time. We implement the entire algorithm in \emph{Mathematica} and compute all the non-trivial connected four-point functions $\langle k_1 k_2 k_3 k_4 \rangle$ with $2 \leq k_1 \leq k_2 \leq k_3 \leq k_4 \leq 8$  ($94$ in total and including $64$ previously unknown correlators), which can be found in the database attached to this publication. Additionally, we compute two very high-weight cases, namely $\langle 7\,10\,12\,17 \rangle$ and $\langle 17 \, 21 \, 23 \, 25 \rangle$. The computation of these latter correlation functions takes $1$ minute and $40$ minutes, respectively, on a standard computer. 

\paragraph{Verification.} We have checked all of these four-point functions except for the high-weight case $\langle 17 \, 21 \, 23 \, 25 \rangle$ for consistency with the structure predicted by superconformal symmetry, see e.g. \cite{Chicherin:2015edu}, using the method described in \cite{Arutyunov:2018neq}. In order to find this structure one should extract the free part of the correlator: in principle the free part can be computed straightforwardly by performing Wick contractions between extended CPOs. However, in our implementation we find the free part from requiring consistency with superconformal symmetry: after reducing the $D$-functions the difference between the four-point function and the prediction from superconformal symmetry splits naturally into four parts as
\begin{equation}
F_{1}\left( \vec{x}, \sigma,\tau \right)+  F_{2}\left( \vec{x}, \sigma,\tau \right) \log u + F_{3}\left( \vec{x}, \sigma,\tau \right) \log v + F_{4}\left( \vec{x}, \sigma,\tau \right) \bar{D}_{1111}, 
\end{equation}
where $u,v$ are the usual conformal cross-ratios
\begin{equation}
u= \frac{x_{12}^2 x_{34}^2}{x_{13}^2 x_{24}^2}, \quad v= \frac{x_{14}^2 x_{23}^2}{x_{13}^2 x_{24}^2}
\end{equation}
and the $F_i$ are rational functions of the $x_i$. The undetermined free part is contained completely in $F_1$ and its vanishing can be used to determine the free part. The vanishing of $F_{2,3,4}$ is independent of the free part and provides a non-trivial check that the computed correlator is indeed consistent with superconformal symmetry.

\paragraph{Mellin representation.}

We can go one step further by checking whether the correlator matches the conjectured formula from \cite{Rastelli:2016nze,Rastelli:2017udc}. In order to do this we first reorder the full correlator $\langle k_1 k_2 k_3 k_4 \rangle$ such that the weights satisfy $k_1 \geq k_2 \geq k_3 \geq k_4$ (which we will distinguish by writing $G_{k_1k_2k_3k_4}$ instead) and then, following \cite{Rastelli:2016nze,Rastelli:2017udc}, write the correlator as
\begin{equation}
G_{k_1k_2k_3k_4}\big(\hspace{.07em}\vec{x}, \vec{t}\,\hspace{.1em}\big) = \mathcal{A}_{k_1k_2k_3k_4}\big(\hspace{.07em}\vec{x}, \vec{t}\, \hspace{.1em}\big) \mathcal{G}_{k_1k_2k_3k_4}\left(u,v,\sigma,\tau \right),
\end{equation}
where the overall prefactor $\mathcal{A}_{k_1k_2k_3k_4}$ depends on the weights as in (4.9) of \cite{Rastelli:2017udc}. One can decompose $\mathcal{G}_{k_1k_2k_3k_4}$ as
\begin{equation}
\label{eq:RHdecomposition}
\mathcal{G}_{k_1k_2k_3k_4} = \mathcal{G}_{k_1k_2k_3k_4}^{\text{free}} + R \mathcal{H},
\end{equation}
where 
\begin{equation}
R = \tau + (1-\sigma -\tau)v + (-\tau - \sigma \tau +\tau^2)u +(\sigma^2 -\sigma -\sigma \tau)uv + \sigma v^2 +\sigma \tau u^2. 
\end{equation}
The conjecture from \cite{Rastelli:2016nze,Rastelli:2017udc} is a simple formula for the dynamical function $\mathcal{H}$ in Mellin space up to an overall weight-dependent normalization, which was later determined in \cite{Aprile:2018efk}. 

Using the free part extracted in the verification process we can easily find an expression for $R \mathcal{H}$. We can furthermore find $\mathcal{H}$ in terms of $\bar{D}$ functions by solving a set of linear equations obtained from the decomposition into different tensor components. This expression can directly be Mellin-transformed and compared to the conjectured formulae and we find agreement for all checked correlation functions: these are all the $91$ (of which $61$ new) correlation functions with weights up to and including $8$ except for the three with lowest weight $k_1 \geq 7$,\footnote{This exception is due to computational limits} as well as $\langle 7 \, 10 \, 12 \, 17 \rangle$. In particular, we find agreement with the derived normalization function from \cite{Aprile:2018efk}, which in our notation becomes
\begin{equation}
\label{eq:normalization}
f(k_1,k_2,k_3,k_4) = \frac{2^4 \sqrt{k_1 k_2 k_3 k_4}}{\left(\tfrac{k_4-k_3+k_2-k_1}{2} \right)!\left(\tfrac{k_4+k_3-k_2-k_1}{2} \right)!\left(\tfrac{|k_4-k_3-k_2+k_1|}{2} \right)!\left( L-2\right)!},
\end{equation}
with $L$ as in \cite{Rastelli:2017udc}:
\begin{equation}
\label{eq:L}
L(k_1,k_2,k_3,k_4) =  \begin{cases}
    k_1       & \quad \text{if } k_1+k_4 \leq k_2 +k_3\\
   \tfrac{k_1+k_2+k_3-k_4}{2}  & \quad \text{otherwise}
  \end{cases}
\end{equation}

\paragraph{Database.}With this publication\footnote{The database can be found in the arXiv submission of this paper.} we include a database of all the non-trivial correlators $\langle k_1 \ldots k_4 \rangle$ with $2 \leq k_1 \leq k_2 \leq k_3 \leq k_4 \leq 8$. Moreover we include the results for the high weight cases $\langle 7\,10\,12\,17 \rangle$ and $\langle 17 \, 21 \, 23 \, 25 \rangle$. For each correlation function there is a subfolder with the name \texttt{$k_1$\_$k_2$\_$k_3$\_$k_4$} containing  up to five plain \texttt{txt} files: 
\begin{itemize}
\item \texttt{Fullcorrelator$k_1$\_$k_2$\_$k_3$\_$k_4$.txt} contains the full correlator as we compute it directly from the action, in the notation from \cite{Arutyunov:2018neq},

\item \texttt{Freepart$k_1$\_$k_2$\_$k_3$\_$k_4$.txt} contains the free part as we extract it from consistency with superconformal symmetry (also in the notation from \cite{Arutyunov:2018neq}),

\item \texttt{H$k_1$\_$k_2$\_$k_3$\_$k_4$.txt} contains a coordinate-space expression for the dynamical function $\mathcal{H}$ in the notation of \cite{Rastelli:2016nze,Rastelli:2017udc} as it follows from our direct computation,\footnote{This expression has not been simplified and is therefore much longer than what is predicted from the Mellin-space conjecture.}

\item \texttt{HfromMellin$k_1$\_$k_2$\_$k_3$\_$k_4$.txt} contains a much shorter coordinate-space expression for $\mathcal{H}$ in the notation of \cite{Rastelli:2016nze,Rastelli:2017udc} and has been derived from its Mellin-space form (the construction of which we discuss in the appendix),

\item \texttt{RZconj$k_1$\_$k_2$\_$k_3$\_$k_4$.txt} contains two entries: if this four-point function coincides with the Mellin conjecture the first entry is \texttt{yes} and if not it would read \texttt{no}. The second entry is the value for the overall scaling function $f(k_1,k_2,k_3,k_4)$. 
\end{itemize}

\section{Conclusion}
\label{sec:conclusions}
In this work we have demonstrated that the simplified algorithm obtained in \cite{Arutyunov:2018neq} together with the harmonic polynomial formalism allow one to compute \textit{any} four-point functions of CPOs of reasonable weights very fast. For example, now the computation of the $\langle 7\,10\,12\,17 \rangle$ correlator takes only a minute. Attached to this publication we provide a database of all the correlators with weights up to and including $8$. As an application of this new simplified algorithm we confirm for most of these correlation functions that they match the Mellin formula from \cite{Rastelli:2016nze,Rastelli:2017udc}, thereby further corroborating this hypothesis. These new correlators go far beyond the previously known set of four-point functions. 

One could use the discussed simplifications to derive a closed formula for the coordinate space correlation function from the supergravity action, which could in turn be used to prove the Mellin conjecture in full generality. The main remaining obstruction sits in the fact that the correlation function cannot at present be written as a closed formula in terms of $D$-functions: the representations appearing in the tensor product decomposition, parametrizing both the couplings and the exchanged fields, depend on the external weights in an intricate way and the exchange Witten diagrams are found algorithmically. 

\section*{Acknowledgments}
We would like to thank Till Bargheer, Vsevolod Chestnov and Enrico Olivucci for useful discussions. In particular, we thank Sergey Frolov for collaborations during the initial stages of this work. This work is supported by the German Science Foundation (DFG) under the Collaborative Research Center (SFB) 676 ``Particles, Strings and the Early Universe" and the Research Training Group (RTG) 1670 ``Mathematics inspired by String Theory and Quantum Field Theory". 

\appendix
\section{Deriving coordinate-space expressions from Mellin space}
\label{app:derivation}
Given a coordinate-space expression for the dynamical function $\mathcal{H}$ of a correlation function 
\begin{equation}
\label{eq:A1}
\sum a_{k_1k_2k_3k_4}(u,v) \bar{D}_{k_1k_2k_3k_4} + \sum b_{k_1k_2k_3k_4}(u,v),
\end{equation}
where the sums are finite and run over the $k_i$ and $a$ and $b$ are rational functions of $u$ and $v$ and we suppress their polynomial dependence on $\sigma$ and $\tau$,
it is straightforward to find the corresponding Mellin-space expression: as discussed in \cite{Rastelli:2017udc} we can consistently send the functions $b$ to zero and use that 
\begin{eqnarray}
\label{eq:Dfunction}
\bar{D}_{\Delta_1 \ldots \Delta_4} (u,v) = 2 \int \frac{ds}{2}\frac{dt}{2} u^{\tfrac{s}{2} - \frac{\Delta_1 +\Delta_2}{2}} v^{\tfrac{t}{2} - \frac{\Delta_2 +\Delta_3}{2}}
\Gamma\left(\tfrac{-s+\Delta_1+\Delta_2}{2} \right)
\Gamma\left(\tfrac{-s+\Delta_3+\Delta_4}{2} \right) \times \\
\Gamma\left(\tfrac{-t+\Delta_1+\Delta_4}{2} \right)
\Gamma\left(\tfrac{-t+\Delta_2+\Delta_3}{2} \right)
\Gamma\left(\tfrac{s+t-\Delta_2-\Delta_4}{2} \right)
\Gamma\left(\tfrac{s+t-\Delta_1-\Delta_3}{2} \right),
	\nonumber
\end{eqnarray}
to find the Mellin transform $\mathcal{M}\left[ \mathcal{H} \right]$ of $\mathcal{H}$, defined as
\begin{eqnarray}
\label{eq:H}
\mathcal{H} = \int \frac{ds}{2} \frac{dt}{2} u^{\tfrac{s}{2}-\tfrac{k_3+k_4}{2} + L} v^{\tfrac{t}{2}-\tfrac{\text{min}\left(k_1+k_4,k_2+k_3 \right)}{2}}\mathcal{M}\left[ \mathcal{H} \right](s,t)\Gamma_{k_1 k_2 k_3 k_4},  
\end{eqnarray}
with $L$ defined in \eqref{eq:L} and 
\begin{equation}
\Gamma_{k_1 k_2 k_3 k_4} = \Gamma\left(\tfrac{-s+k_1+k_2}{2} \right)
\Gamma\left(\tfrac{-s+k_3+k_4}{2} \right)
\Gamma\left(\tfrac{-t+k_1+k_4}{2} \right)
\Gamma\left(\tfrac{-t+k_2+k_3}{2} \right)
\Gamma\left(\tfrac{s+t+4-k_2-k_4}{2} \right)
\Gamma\left(\tfrac{s+t+4-k_1-k_3}{2} \right).
\end{equation}
However, finding an expression of the form \eqref{eq:A1} by inverse-Mellin transforming $\mathcal{M}\left[ \mathcal{H} \right]$ is not a straightforward task, but it can be done. To see how one can do this we only need to consider a single summand in the Mellin representation of a correlator: so let us for simplicity assume
\begin{equation}
\mathcal{M}\left[ \mathcal{H} \right]\sim \frac{1}{(s-s_0)(t-t_0)(\tilde{u}-\tilde{u}_0)},
\end{equation}
with $\tilde{u} = k_1 +k_2 +k_3 +k_4 -4 -s -t$ and $s_0,t_0,\tilde{u}_0$ non-negative integers. In order to inverse Mellin-transform such an expression we will consider part of the integrand in \eqref{eq:H}, namely
\begin{equation}
\label{eq:integrand}
\frac{\Gamma_{k_1 k_2 k_3 k_4}}{(s-s_0)(t-t_0)(\tilde{u}-\tilde{u}_0)}.
\end{equation}
If we manage to rewrite this expression as a linear combination of the $\Gamma_{p_1 p_2 p_3 p_4}$, each summand in that sum gives rise to an integral of the form \eqref{eq:Dfunction} after an appropriate identification of the $\Delta_i$ with the $p_i$. Therefore, the problem of finding a coordinate-space expression is reduced to finding a linear combination of $\Gamma_{p_1 p_2 p_3 p_4}$ such that
\begin{equation}
\label{eq:csum}
\frac{\Gamma_{k_1 k_2 k_3 k_4}}{(s-s_0)(t-t_0)(\tilde{u}-\tilde{u}_0)} = \sum c_{p_1 p_2 p_3 p_4} \Gamma_{p_1 p_2 p_3 p_4},
\end{equation}
where the $c$ are numbers and the sum is finite over the $p_i$. The representation on the right-hand side of \eqref{eq:csum} is usually not unique, which reflects the fact that the $\bar{D}$ functions are not independent. 

The first step towards an expression as in the right-hand side of \eqref{eq:csum} is to rewrite its left-hand side as a pure product of gamma functions and linear factors. This can be done by applying the basic property $x \Gamma(x) = \Gamma(x+1)$ repeatedly to some of the gamma functions in the numerator, such that finally factors in the numerator cancel the denominator. For example 
\begin{equation}
\frac{\Gamma(\frac{-s+5}{2})}{(s-1)} = \frac{\frac{-s+3}{2}\frac{-s+1}{2}\Gamma(\frac{-s+1}{2})}{(s-1)} = -\frac{1}{2}\frac{-s+3}{2}\Gamma\left(\frac{-s+1}{2}\right).
\end{equation}
This yields the intermediate form
\begin{equation}
\label{eq:factors}
C \frac{-s+s_1}{2}\ldots \frac{-s+s_n}{2}\frac{-t+t_1}{2} \ldots \frac{-t+t_m}{2}\ldots \frac{-\tilde{u}+\tilde{u}_1}{2}\ldots \frac{-\tilde{u}+\tilde{u}_l}{2} \Gamma(x_1)\ldots \Gamma(x_6),
\end{equation}
with $C$ some constant and $x_{1,2}$ a linear factor in $s$, $x_{3,4}$ a linear factor in $t$ and $x_{5,6}$ a linear factor in $\tilde{u}$. Note that it could happen that each of the three sets of prefactors in $s$, $t$ and $\tilde{u}$ might be empty. Suppose first for simplicity that there is only one factor $\frac{-s+s_1}{2}$. Let us consider the linear equation
\begin{equation}
\frac{-s+s_1}{2} = \sum_{i=1}^6 \lambda_i x_i
\end{equation}
for the six unknowns $\lambda_i$. Working out the arguments $x_i$ one sees that there are four independent equations (one for $s$, $t$ and $\tilde{u}$ and one for the constant part), such that we are guaranteed a solution. With this solution we can now rewrite 
\begin{equation}
\frac{-s+s_1}{2} \Gamma(x_1)\ldots \Gamma(x_6)  = \sum_{i=1}^6 \lambda_i x_i\Gamma(x_1)\ldots \Gamma(x_6) = \sum_{i=1}^6 \lambda_i \Gamma(x_1)\ldots\Gamma(x_i+1)\ldots \Gamma(x_6)
\end{equation}
and see that we have succeeded in our goal: by repeating the procedure described above recursively for the list of factors in \eqref{eq:factors} we can find a linear combination of products of gamma functions that are equal to \eqref{eq:integrand}, such that we have found a representation as in \eqref{eq:csum}. Exchanging sum and integral we find that each summand is of the form \eqref{eq:Dfunction} such that after matching the coefficients we find an expression for the inverse Mellin-transform in terms of $\bar{D}$ functions. 

We have applied this algorithm to all the correlation functions in our database and included the result in the database. All cases have been checked explicitly with our coordinate-space results. In some cases it is possible that a more minimal representation exists, but due to the automatized nature of our application this is unavoidable. It is noteworthy that in exchanging the sum and integral during this procedure we do not run into any domain issues that exist for the full correlator as described in \cite{Rastelli:2017udc} that give rise to the free part upon inverse Mellin-transforming. After all, all we are doing is rewriting the integrand using a global property of the gamma function. 
\bibliography{bibliography}

\end{document}